\documentclass[apl,twocolumn,showpacs,amsmath,letter]{revtex4}
\usepackage{graphicx}

\newcommand{\Journal}[4]{#1 \textbf{#2}, #3 (#4)}

\begin{document}

\title{Effect of Antiferromagnetic Interlayer Coupling on Current-Assisted Magnetization Switching}

\author{S. Urazhdin}
\author{H. Kurt}
\author{W. P. Pratt Jr.}
\author{J. Bass}
\affiliation{Department of Physics and Astronomy, Center for Fundamental Materials Research,
and Center for Sensor Materials, Michigan State University, East Lansing, MI 48824-2320}


\begin{abstract}
We compare magnetization switching in Co/Cu/Co nanopillars with
uncoupled and dipole-field coupled Co layers.  In uncoupled
nanopillars, current-driven switching is hysteretic at low
magnetic field H and changes to reversible, characterized by
telegraph noise, at high H.  We show that dipolar coupling both
affects the switching current and causes the switching to become
reversible at small H. The coupling thus changes the switching to reversible,
hysteretic, and then reversible again as H increases.  We describe
our results in terms of current-assisted thermal activation.
\end{abstract}

\maketitle

Observations of current-driven magnetization direction
switching in Co/Cu/Co~\cite{cornellorig,cornellapl,grollier,cornelltemp,cornellquant,wegrowe,sun2}
and Py/Cu/Py [Py = Permalloy = Ni$_{84}$Fe$_{16}$]~\cite{urazhdin}
nanopillars have generated great interest, both for
science---studies of magnetic systems driven far out of
equilibrium, and technology---in magnetic
random access memory (MRAM) this effect might eliminate the need for
magnetic field-driven switching.
In this Letter we show that dipolar coupling between the
magnetic layers can affect the switching. Specifically,
we show that the switching current is not determined solely by the
switching mechanism~\cite{cornellquant, grollier}, but varies with
dipolar coupling, and that sufficiently strong antiferromagnetic
(AF) coupling leads to reversible (non-hysteretic) switching at
small magnetic field H. This effect may find application in
high-sensitivity field sensors, e.g. read-heads of computer hard
drives.

Our devices were nanofabricated using the
following steps.  First, a
Cu(80)/Co(20)/Cu(6-10)/Co(2.5)/Cu(5)/Au(15) multilayer was
sputtered onto an oxidized Si wafer in Ar pressure of 2~mTorr. All
thicknesses in this Letter are in nanometers.  An Al(50) nanopillar with lateral dimensions of
about 70 nm by 130 nm was then formed by a combination of e-beam lithography
and evaporation.  The Al was used as a mask for ion-milling
the multilayer into a nanopillar. Dipolar coupling between the two Co
layers was minimized by timing the ion-milling to stop in the
Cu layer above the bottom Co(20) layer. When desired, AF
dipolar coupling was achieved by continuing the ion-milling about
half way through the bottom Co(20) layer.  Magnetic poles at the
edges of the two patterned Co layers then AF-couple
to minimize the magnetic energy.  Without
breaking the vacuum, a SiO(25-40) layer was deposited for electrical
insulation between the device electrodes.  The Al mask was removed
by ion-milling with the ion-beam parallel to the
sample surface, followed by wet etching.  The ion-milling removed
metals back-sputtered onto the Al mask.
Finally, after a short ion-milling to clean the surface, a Au(150)
top contact was sputtered onto the top Au layer.  All measurements
were performed at room temperature (295~K).  Differential
resistances dV/dI were measured using a standard four-probe setup
with lock-in detection, adding an ac current of amplitude 20~$\mu$A at
8~kHz to the dc current I. Most uncoupled devices tested had
room temperature resistances of about 1.5~$\Omega$ and magnetoresistances
(MR) of about 5\%, similar to the best MR values reported by
others~\cite{cornellorig}. The coupled devices had
larger resistances (because of the additional interface and Co
layer in the nanopillar) but similar MRs. We define positive
current to flow from the bottom (extended) to the top Co layer.

\begin{figure}
\includegraphics[scale=0.41]{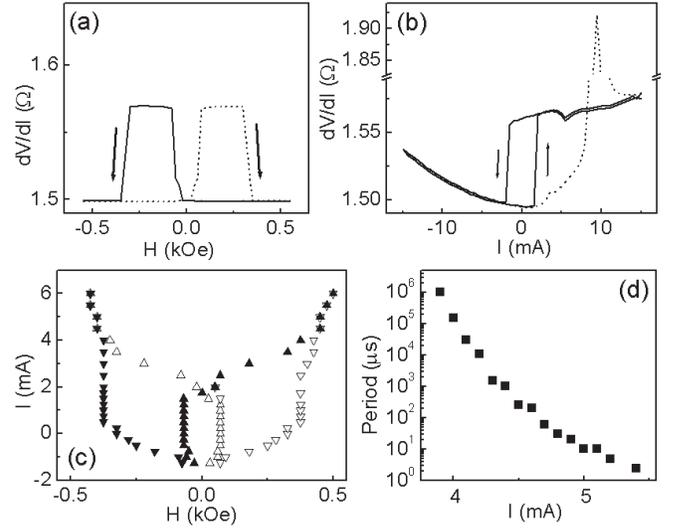}
\caption{\label{fig1} Magnetization switching in an uncoupled sample.
(a) H-dependence of dV/dI at I=0. (b) Current dependence of dV/dI:
Solid line: H=0~Oe, dashed line: H=600~Oe. Arrows mark the scan
direction. (c) Switching diagram extracted from the H scans at
various fixed values of I. Upward triangles: P$\to$AP switching, as
defined in the text. Downward triangles: AP$\to$P switching. Open
symbols: H scanned from negative to positive, closed
symbols: reverse H-scan. (d) Variation
of the average telegraph noise period with I, as H is varied
between 426~Oe and 479~Oe to keep $\tau_P=\tau_{AP}$.}
\end{figure}

\begin{figure}
\includegraphics[scale=0.41]{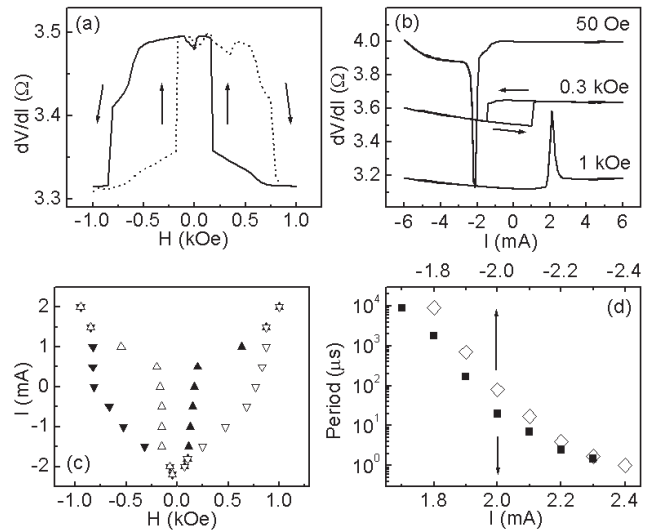}
\caption{\label{fig2} Magnetization switching in a dipole-coupled
sample. (a) H dependence of dV/dI at I=0. (b) Current dependence of dV/dI
at various H. Curves are offset for clarity. Arrows mark
the scan direction at H=0.3~kOe. (c)
Switching diagram extracted from the H scans at various fixed
values of I. Upward triangles: P$\to$AP switching, as defined in
the text. Downward triangles: AP$\to$P switching. Open symbols: H
scanned from negative to positive, closed symbols: reverse H-scan.
(d) Variation of the average
telegraph noise period with I, as H is varied to keep
$\tau_P=\tau_{AP}$. Solid symbols: $I>0$ (bottom scale), H varied
between 854 and 968~Oe. Open symbols: $I<0$ (top scale), H
varied between 57 and 128~Oe.}
\end{figure}

Typical data for an uncoupled sample are shown in Fig.~\ref{fig1},
and comparative data for a coupled sample are shown in
Fig.~\ref{fig2}. We begin with the uncoupled sample, the data for
which are consistent with prior studies of uncoupled
Co/Cu/Co~\cite{cornellapl} and Py/Cu/Py~\cite{urazhdin}.
Fig.~\ref{fig1}(a) shows that the MR at I = 0 is symmetrically
hysteretic in H and the increase in resistance from the low
resistance parallel (P) orientation of magnetic moments to the
high resistance anti-parallel (AP) orientation occurs in a single
sharp step. The current-driven switching is hysteretic at H = 0~Oe
(solid curve in Fig.~\ref{fig1}(b)), with switching to the high
resistance anti-parallel (AP) state at positive $I_s^{P\to AP}$,
and to the low resistance parallel (P) state at negative
$I_s^{AP\to P}$. The resistances and their changes in
Fig.~\ref{fig1}(b) are close to those in Fig.~\ref{fig1}(a).  The
dotted curve in Fig.~\ref{fig1}(b) shows that at large enough H
the hysteretic step in dV/dI turns into a non-hysteretic
(reversible) switching peak.

Fig.~\ref{fig1}(c) shows the switching diagram obtained from
MR data such as those in Fig.~\ref{fig1}(a). We attribute the
slight H-asymmetry of the diagram to a combination of the
current-induced Oersted field and sample shape asymmetry. The P$\to$AP
switching field $H_s^{P\to AP}\approx70$~Oe is independent of I
over the range $-1.7$~mA$<I<1.7$~mA.  We attribute the P$\to$AP
switching in this range to reversal of the extended Co layer,
unaffected by the small current density in that layer.  We
associate the AP$\to$P transition with reversal of the thin
patterned Co layer at $H_s^{AP\to P}$ determined by its shape
anisotropy.  At $I>1.7$~mA, $H_s^{P\to AP}$ strongly varies with
I. In this regime, we attribute the  P$\to$AP transition to
reversal of the patterned Co layer, induced by $I>0$.  As the
extended Co layer reverses at $H\approx70$~Oe, the patterned Co
layer reverses simultaneously to keep the AP configuration favored
by $I>0$. Such simultaneous reversal sometimes produces a weak
feature in dV/dI at small H.

At $I >4$~mA, the data in Fig.~\ref{fig1}(c) become
non-hysteretic. Time resolved measurements show that the
non-hysteretic switching peak in dV/dI is characterized by
telegraph noise switching between the P and AP states, with random
distribution of dwell times in the P state ($\tau_P$) and AP state
($\tau_{AP}$)~\cite{urazhdin}.
 When I is increased, and H adjusted to keep
$\tau_P=\tau_{AP}$, the average noise period decreases
approximately exponentially (Fig.~\ref{fig1}(d)). The switching diagram,
Fig.~\ref{fig1}(c), is asymmetric with respect to the current
direction; switching is hysteretic at $I<0$, and
non-hysteretic only at large $I>0$. This difference occurs because
application of H favors the P state, while $I>0$ favors the AP
state.

Fig.~\ref{fig2} shows data similar to those in Fig.~\ref{fig1},
but for a sample with strong AF dipolar coupling between the two
Co layers, produced by partial patterning of the extended Co(20)
layer. Again, the field-driven MR (Fig.~\ref{fig2}(a)) is
hysteretic and approximately symmetrically about $H = 0$.  But now the
MR contains multiple steps, likely because the highly inhomogeneous
dipole field favors nonuniform magnetization states of the
nanopillar. We determine the switching points from the jump
into or from the lowest resistance state. As H is reduced from a
large positive value (solid line in Fig.~\ref{fig2}(a)), coupling
between the two Co layers causes the thinner Co (2.5) layer to
switch to the AP state at $H=0.16$~kOe, then at small $H<0$ both
the thicker and thinner  Co layers flip together to stay in the AP
state, and finally at $H\approx -0.8$~kOe the Co(2.5) layer
reverses again to return to the P state.

At small H =50~Oe the behavior as a function of I in
Fig.~\ref{fig2}(b) is quite different from that in
Fig.~\ref{fig1}(b); instead of hysteretic switching, the data show
a non-hysteretic, downward peak.   This behavior is the most
important feature of the data for coupled samples.
In Fig.~\ref{fig2}(b), the switching is non-hysteretic for small H, becomes hysteretic
for intermediate H=0.3~kOe, and then non-hysteretic again for large enough H.
Fig.~\ref{fig2}(c) shows the switching diagram.
In contrast to Fig.~\ref{fig1}(c), all switching points
now represent reversal of the the Co(2.5) layer and switching
becomes reversible at large enough magnitude of current in both
directions, at small H for negative I, and large H for positive I.
The symmetry between effects of positive and negative
current is extended further by time-resolved measurements of
telegraph noise, which is present close to the reversible
switching points for both current directions. Fig. 2(d) shows that
as the magnitude of I is increased, while adjusting H to keep
$\tau_P=\tau_{AP}$, the telegraph noise period decreases
approximately exponentially at similar rates for both current directions.
Not all samples with partially patterned bottom Co layer exhibited non-hysteretic switching at small H.
Due to weaker AF-coupling, some samples exhibited only a dip in $I_s^{P\to AP}$ at small H, resulting
in switching diagrams intermediate between Fig.~\ref{fig1}(c) and Fig.~\ref{fig2}(c).

\begin{figure}
\includegraphics[scale=0.35]{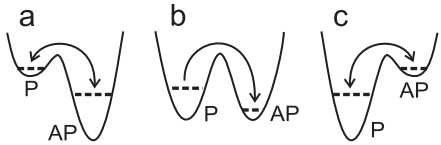}
\caption{\label{fig3} (a) Schematic of current-driven telegraph
noise in AF-coupled sample at $I<0$, small H. Dashed lines
indicate $T_m$. (b) Same as (a), at $I>0$, H close to the dipole field.
(c) Same as (a), at $I>0$, large H.}
\end{figure}

We interpret the data in Figs.~\ref{fig1},\ref{fig2} in terms of thermal activation over an
effective switching barrier~\cite{wegrowe,cornelltemp,urazhdin}.
In Co/Cu/Co nanopillars, both the
semi-classical spin-torque~\cite{slonczewski} and
quantum threshold~\cite{berger,tsoi} models predict magnetic
excitation in the P state at $I>0$ and in the AP state at $I<0$.
Such current-driven excitation can be described in terms of a current-dependent effective
switching barrier~\cite{zhang}, or a current-dependent effective magnetic temperature
$T_m$, with an effective barrier modified by the current only through the variation of magnetization
with temperature~\cite{wegrowe,urazhdin}.  Here $T_m$ depends on the magnetic configuration,
$T_m^{AP}(I)\not= T_m^P(I)$. These alternative approaches give mathematically similar
results, but differ in details that still need experimental testing. We use the latter model,
which we find more transparent.

In magnetically uncoupled samples, the magnetization orientation
of the patterned layer is bistable at $H=0$, $I=0$: at T=295~K,
the barrier height for switching between the two magnetization
orientations significantly exceeds the thermal energy. In
AF coupled samples at small H, the P$\to$AP switching
barrier is reduced by the dipolar field, leading to a thermally
activated P$\to$AP transition. The reverse AP$\to$P
transition cannot be thermally activated because the corresponding
switching barrier is significantly higher. Thus, at I=0, H=0, AP
is the only stable orientation of the nanopillar, as seen in
Fig.~\ref{fig2}(b,c). Current-driven magnetic excitation
at $I<0$ increases $T_m^{AP}$. At sufficiently high $I<0$ the AP$\to$P transition
becomes thermally activated, leading to telegraph noise switching
between the AP and P states (Fig.~\ref{fig3}(a)). At larger
H=0.3~kOe in Fig.~\ref{fig2}(b), H nearly compensates the dipole
field, leading to hysteretic switching similar to that in
Fig.~\ref{fig1}(a) at H=0 in an uncoupled sample, and illustrated
in Fig.~\ref{fig3}(b). As H is further increased, both the
uncoupled and coupled samples behave similarly; As shown in
Fig.~\ref{fig3}(c), the AP$\to$P transition becomes thermally
activated. At large enough $I>0$ the P$\to$AP transition
also becomes activated due to current-induced increase
of $T^P_m$, leading to telegraph noise both in uncoupled
(Fig.~\ref{fig1}(d)) and AF-coupled (Fig.~\ref{fig2}(d))
nanopillars.

In summary, we have shown that dipolar AF coupling between
magnetic layers leads to reversible current-driven magnetization
switching at small H. Similarly, reversible switching at small H
should be induced by the opposite current direction in
ferromagnetically exchanged-coupled samples. We will demonstrate such behavior
elsewhere~\cite{urazhdin2}. Thus, the switching current and hysteresis are not intrinsic
characteristics of the current-driven switching mechanisms, as
they are strongly affected by the coupling between the magnetic
layers.

Fig.~\ref{fig2}(c) shows that, in the reversible switching regime at small H,
the switching field of AF-coupled magnetically nanopillars can be
adjusted by changing the applied current. In this regime, magnetically coupled nanopillars
with exchange-biased extended magnetic layer and zeroed-out switching field
may find application as high sensitivity field sensors.

We acknowledge helpful discussions with Norman O. Birge, important
contributions to the development of the sample preparation technique by
K. Eid and J. Caballero, and support from the MSU CFMR, CSM, the MSU Keck
Microfabrication facility, the NSF through Grants DMR 02-02476 and
98-09688, and Seagate Technology.

\end{document}